**Spontaneous rotation of magnets levitating above high-$T_c$ superconductor**

D. M. Gokhfeld[1,2,*], S. Yu. Shalomov[3], D. B. Sultimov[2], M. I. Petrov[1]

1 Kirensky Institute of Physics, Krasnoyarsk Scientific Center, Siberian Branch, Russian Academy of Sciences, 660036 Krasnoyarsk, Russia

2 Krasnoyarsk State Pedagogical University, 660049 Krasnoyarsk, Russia

3 Phys.-Math. School, Siberian Federal University, 660041 Krasnoyarsk, Russia

* e-mail: gokhfeld@iph.krasn.ru

The levitation of a cylindrical permanent magnet over a high-temperature superconductor cooled by liquid nitrogen can be accompanied by spontaneous oscillations and rotation. The reason for spontaneous rotation of the magnet is magnetization inhomogeneity induced by the temperature gradient. An experiment was carried out on the levitation of Nd–Fe–B magnets over a composite high-temperature superconductor. The experimental results confirm that the rotation frequency depends on the difference in the magnetization values in the upper and lower halves of the magnet. Methods for controlling the rotation frequency of a levitating magnet are proposed.

**1. Introduction**

The discovery of high-$T_c$ superconductors with a critical temperature $T_c$ above the boiling point of nitrogen has led to a significant increase in the use of superconducting devices in electrical engineering and energy. One of the consequences of this discovery was the possibility of demonstrating the shielding of the magnetic field by a superconductor and magnetic levitation not only in laboratories [1] but also in classrooms [2,3]. In the late 1980s, it was noticed that levitating cylindrical magnets (mainly Nd-Fe–B) under certain conditions begin to spontaneously swing and rotate (for example, see section 5.1 in the book [4]). Studies of the spontaneous rotation of magnets [5–7] have revealed the following patterns. The rotation frequency is related to the temperature gradient along the vertical diameter of the magnet. The inhomogeneities in the azimuthal magnetization and the deviation of the magnetic axis from the center of mass slow the rotation. The rotation frequency decreases with decreasing distance from the magnet to the surface of the liquid nitrogen. Heating the magnet with infrared radiation leads to an increase in the rotation speed [6].

To explain the observations, the authors of [5–7] used the following experimental data: 1) The magnetic moment of the magnet material depends on the temperature. 2) There is a temperature gradient along the vertical axis. 3) If the magnetization of the cold part of the magnet is greater than that of the warm part, then the average point of action of the electromagnetic repulsion forces becomes lower than the center of mass of the magnet. Exceeding the center of mass above the average point of action of the electromagnetic repulsion forces leads to spontaneous swinging and rotation. The rotation frequency depends on the thermal conductivity of the magnet material and the dynamic temperature difference between the upper and lower halves of the magnet. Thus, a heat engine is realized, which performs work due to the temperature difference between liquid nitrogen and the room. A detailed mathematical analysis of the rotation of the levitating magnet was carried out in [8-10]. Although all the experimental observations fit into this concept, additional research with other types of superconductors and magnets is desirable. Possible practical applications of this phenomenon require knowledge of ways to increase and decrease the frequency of rotation of the magnet. In 2002, one of us (M.I.P.) carried out observations of the spontaneous rotation of magnets levitating over polycrystalline HTSs of various compositions. In the present paper, we investigated the spontaneous rotation of cylindrical magnets using a platform made of layers of composite superconducting tape. At the same time, we were interested in checking the previously proposed explanation of spontaneous rotation, as well as ways to change the frequency of rotation of magnets.

**2. Materials and methods**

Cylindrical Nd–Fe–B magnets from three different sets (designated I, II, III) were used. The dimensions and weights of the magnets are shown in Table 1. All magnets are magnetized along the main axis of rotation. The levitation of the magnets was achieved using a homemade superconducting platform (Figure 1a). The platform is assembled from nine twelve-millimeter composite superconducting tapes (REBCO) produced by SuperOx [11]. The tapes are fastened together in three layers of three in a row.

Video recordings of the levitation and rotation were made for all magnets. After establishing a constant rotation speed, the number of revolutions of the magnet over a period of time was calculated, and the rotation frequency was determined.

A heat gun was used to heat the magnets. Weak pressure was applied so as not to move the magnet. The temperature of the exhaust air jet is 150 °C.

The temperature above the surface of the liquid nitrogen was measured using a Cernox RTD CX-1050 thermal sensor.

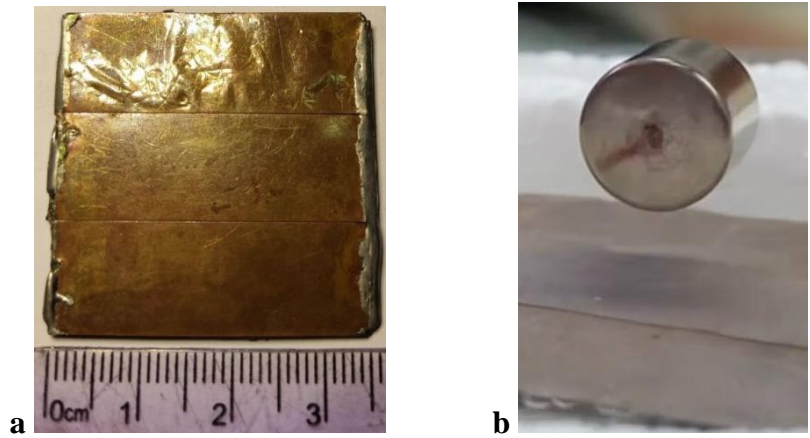
**Fig. 1.** (a) Platform from superconducting tapes. (b) Levitating magnet with a diameter of 10 mm.

**Table 1.** Sizes, masses, and rotation frequencies of Nd-Fe-B magnets.

|     | № | Diameter, mm | Width, mm | Mass, g | Frequency, sec$^{-1}$ |
|-----|---|--------------|-----------|---------|-----------------------|
| I   | 1 | 5.0          | 1.8       | 0.25    | 0                     |
|     | 2 | 10.0         | 10.0      | 6.1     | 2.2                   |
|     | 3 | 12.0         | 2.8       | 2.3     | 2.1                   |
|     | 4 | 20.0         | 3.0       | 7.5     | 1.8                   |
| II  | 5 | 11.9         | 10.0      | 8.3     | 1.6                   |
|     | 6 | 15.0         | 5.0       | 6.7     | 1.5                   |
|     | 7 | 20.0         | 5.0       | 11.8    | 1.1                   |
|     | 8 | 20.1*        | 3.0       | 5.3     | 1.1                   |
|     | 9 | 25.0         | 2.9       | 10.7    | 0.6                   |
| III | 10| 9.0          | 6.4       | 3.0     | 1.2                   |

*Hollow cylinder with a hole diameter of 10 mm.

## 3. Results

### 3.1. Rotation frequency for different magnets

The magnets levitate at a height of approximately 5 mm above a superconducting platform cooled with liquid nitrogen (Figure 1b). After ~ 1-10 s, spontaneous swinging of the magnet begins; that is, the magnet begins to turn relative to the horizontal axis of rotation of the cylinder by several degrees clockwise and counterclockwise. The turn angle gradually increases until it exceeds 180°, after which the magnet begins to rotate. Swinging before the start of rotation lasts less than 1 minute. Furthermore, the rotation accelerates, and the rotation frequency reaches a constant value after ~ 1 minute. Spontaneous rotation was observed for all magnets except for the smallest magnet with a diameter of 5 mm, whose undamping oscillations did not turn into rotation. According to available observations, the direction of rotation of the magnet is set randomly each time. However, the direction of rotation can be set by twisting the magnet during swinging.

The measured rotation frequencies of various magnets are shown in Table 1. There is no general pattern between the size or mass of all magnets and their rotation frequency. However, in sets I and II, a monotonic decrease in the rotation frequency occurs with an increase in the diameter of the magnets (Fig. 2a). For magnets from set II, a decrease in the rotation frequency is observed with an increase in the moment of inertia (Fig. 2b). Additionally, in sets I and II, a monotonic decrease in the rotation frequency occurs with an increase in the temperature difference between the upper and lower points of the magnets ΔT (Fig. 2c).

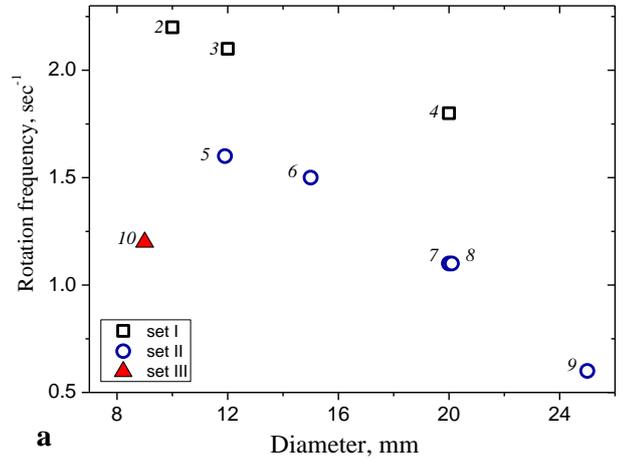

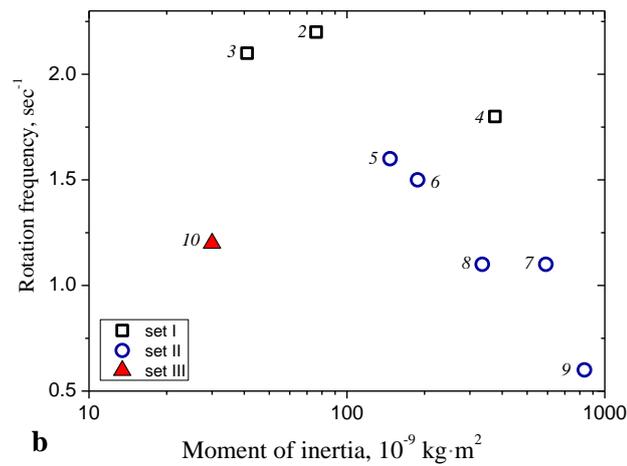

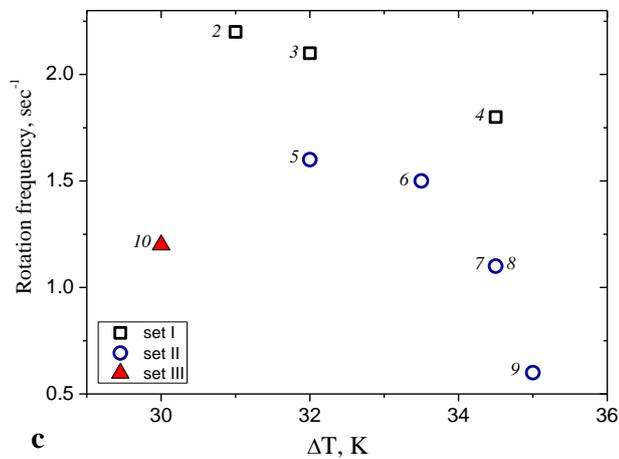

**Fig. 2.** The frequency of rotation of magnets depends on their (a) diameter, (b) moment of inertia and (c) temperature difference between the upper and lower points ΔT. The numbers in the figures indicate the serial number of the magnets in Table 1.

*3.2. Rotation frequency for different levels of liquid nitrogen*

The behavior of levitating magnets with increasing and decreasing levels of liquid nitrogen was considered.

a) Increasing the nitrogen level. At the beginning of the measurements, the nitrogen level was slightly greater than that of the platform, and the lower point of the levitating magnet was approximately 5 mm from the nitrogen level; this is the usual mode for which the rotational speed values are obtained, as shown in Table 1.

As the nitrogen level increased, the rotation of the magnet gradually slowed. The rotation stopped when the lower part of the magnet was immersed in liquid nitrogen. With a further increase in the nitrogen level, there was no rotation or swinging of the magnet.

b) Lowering the nitrogen level. The magnet was placed above the platform at a nitrogen level above the top point of the magnet. At the same time, the magnet does not start to swing, and there is no rotation. Next, we observed the behavior of the magnet during evaporation and a gradual decrease in the level of liquid nitrogen. The vibrations of the magnet did not begin even when the nitrogen was lowered to the level at which the rotation of the magnet was previously observed. To start swinging and rotating, the magnet had to warm above the nitrogen surface for several minutes.

*3.3. Rotation frequency influenced by heat supply*

A wide jet of hot air directed to the center of the magnet (along the axis of rotation) led to an increase in the rotation frequency by more than 2 times compared to the data in Table 1. We assumed that local heating at different points of the magnet can regulate the rotation frequency. Local heating was carried out in the upper or lower halves of the rotating magnet using a narrow nozzle on an air gun. With local heating in the upper region, the rotation frequency also increased 2-2.5 times. However, with local heating of the magnet in the lower region, the rotation frequency decreased until the rotation stopped completely. When the heating stopped, the rotation resumed, reaching the values shown in Table 1.

**4. Discussion**

The rotation of the magnet is caused by the difference in magnetization between the lower and upper halves of the magnet, $M_{down}$ and $M_{up}$. The greater this difference $\Delta M = M_{down} - M_{up}$, the greater the torque. However, $\Delta M$ is not a monotonic function of the temperature gradient. The magnetization of Nd–Fe–B changes nonmonotonically with temperature (Figure 3a) [5,6]. For the magnets used in this work, the dependences of M(T) may have quantitative deviations from the curve shown in Figure 3a. At the same time, the shown dependence is qualitatively fulfilled for all Nd–Fe–B composite magnets.

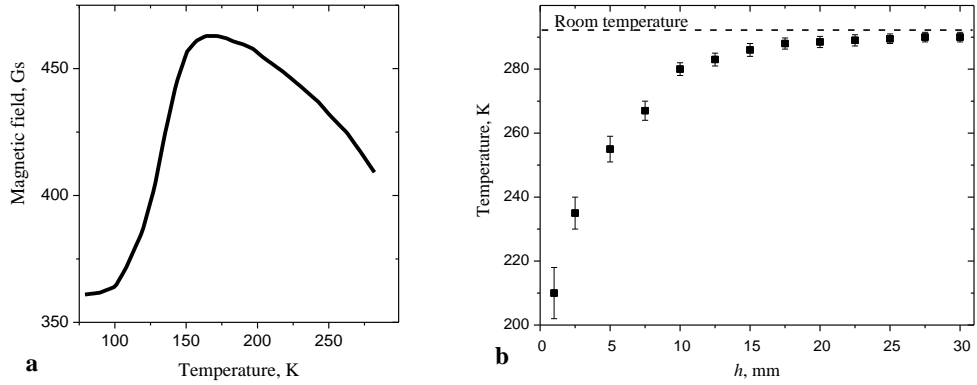

**Fig. 3.** (a) The temperature dependence of the field near the pole of a cylindrical Nd–Fe–B magnet from [6]. (b) Temperature change with increasing height h from the surface of liquid nitrogen.

Figure 3b shows the temperature measured by a thermal sensor at different heights from the surface of liquid nitrogen. From the obtained temperature dependence on the height, it follows that in normal operation, when the magnet is at a height of approximately 5 mm above the liquid nitrogen level, the temperature difference between the upper and lower points for magnets No. 2-10 is 30-35 K. In this case, the temperatures at the upper and lower points of the magnet fall into the temperature range of 160-300 K, where the magnetization of Nd–Fe–B decreases with increasing temperature. Therefore, at a sufficient distance from the liquid nitrogen level, the magnetization of the cold lower half of the magnet is greater than the magnetization of the warmer upper half. Local heating of the lower region reduces the value of $\Delta M$ and slows rotation. Local heating of the upper region leads to an increase in the value of $\Delta M$ and acceleration of rotation.

As the nitrogen level increases, the lower part of the magnet begins to cool, and $M_{down}$ decreases since in the range below 160 K, the magnetization of Nd–Fe–B decreases with decreasing temperature (Figure 3a). When $M_{down}$ begins to be about $M_{up}$ or becomes smaller, the rotation of the magnet stops.

The absence of rotation during levitation of a cylindrical magnet with a diameter of 5 mm is probably caused by a small temperature gradient along the vertical diameter and rapid heat exchange in this magnet.

We also observed a decrease in the rotation frequency of the magnets when their axis of rotation was tilted. This decrease can be explained by the fact that the temperature difference between the upper and lower halves of the magnet decreases when the magnet is tilted [7].

The long swinging time indicates the heterogeneity of the samples [7] or the heterogeneity of their azimuthal magnetization [6]. Indeed, we observed an increase in the swinging time for samples that received small chips at the edges.

It follows from the experiments carried out that the rotation speed is related to the spatial distribution of the magnetization of a cylindrical magnet. This distribution is established due to the temperature gradient and thermal conductivity of the magnet material. One would expect that an increase in the temperature gradient along the vertical axis of the magnet would lead to an increase in the rotational speed due to an increase in ΔM. However, the rotation frequency decreased with increasing diameter of the magnets and ΔT (Figure 2a,c). It should be noted that ΔT, and consequently ΔM, increases slowly with increasing diameter of magnets (see Figure 3b). At the same time, the moment of inertia increases rapidly with increasing sample diameter. Consequently, the effect of slowing rotation due to an increase in the moment of inertia is more significant than the effect of an increase in the temperature gradient.

## 5. Conclusion

The spontaneous rotation of cylindrical Nd–Fe–B magnets during their levitation over composite superconducting tapes is investigated. This phenomenon serves as a good demonstration of the strong diamagnetism of a superconductor and the operation of a heat engine. These observations confirm that the cause of rotation is the temperature gradient and the resulting inhomogeneity of the magnetization of the magnets. It has been found that it is possible to control the rotation frequency of a magnet by changing the nitrogen level or locally heating the magnet. Heating the upper half accelerates, and heating the lower half slows the rotation of the magnets.

**Acknowledgments**

The authors would like to thank A.L. Friedman for creating the superconducting platform, S.V. Semenov for his help with temperature measurements, and S.V. Komogortsev for useful discussions. The study is performed within the state assignment of Kirensky Institute of Physics.